\documentclass[runningheads]{llncs}
\usepackage[T1]{fontenc}
\usepackage{amsmath, amssymb, bbm}
\usepackage[misc]{ifsym}
\usepackage{url}
\usepackage{mwe}

\usepackage{tabularx}
\usepackage{multirow}
\usepackage{makecell}
\usepackage{adjustbox}
\usepackage{graphicx} 
\usepackage{booktabs} 
\usepackage{float}
\usepackage[table]{xcolor}

\newcommand{\rapt}{\textsc{Rapt}\xspace}
\newcommand{\bhalf}{Base$_{0.5}$}
\newcommand{\bopt}{Base$^{*}$}
\newcommand{\bloc}{Base$^{*}_{l}$}
\newcommand{\cav}{\textsc{Rapt}$_{\#}$}
\newcommand{\crk}{\textsc{Rapt}$_{\geq}$}
\newcommand{\companyname}{WM Nicol\xspace}

\begin{document}

\title{RAPT: Retrieval-Augmented Post-hoc Thresholding for 
Multi-Label 
Classification}
\titlerunning{RAPT for 
Multi-Label 
Classification}

\author{
Lasal Jayawardena\inst{1} \and
Nirmalie Wiratunga\inst{1} \and
Ikechukwu Nkisi-Orji\inst{1} \and
Darren Nicol\inst{2}
}

\authorrunning{L. Jayawardena et al.}

\institute{
Robert Gordon University, Aberdeen, United Kingdom \\
\email{\{l.jayawardena,n.wiratunga,i.o.nkisi-orji\}@rgu.ac.uk}
\and
William Nicol (Aberdeen) Limited, Aberdeen, United Kingdom \\
\email{darren.nicol@williamnicol.co.uk}
}

\maketitle              

\begin{abstract}
Industrial multi-label document understanding pipelines score candidate labels and threshold or rank them to form a label set per document.
This early selection step directly affects the accuracy of downstream information extraction from the document, as well as the associated verification effort.
In practice, OCR noise, label imbalance, instance-dependent label cardinality, and asymmetric error costs make global score thresholds brittle and hard to maintain as document formats evolve. 
We present \rapt, a deployment-oriented retrieval-augmented score thresholding wrapper, applied post-hoc to improve label set selection without retraining the underlying classifier. 
\rapt is a model-agnostic wrapper: any predictor that provides document representations for similarity search and per label confidence scores can be used, including metric learning encoders and fine-tuned transformer classifiers. 
For each query document, given a classifier’s score vector, \rapt retrieves similar document thresholding situations (cases) and adapts the query's label set selection threshold using their outcomes.
The adaptation selects the final label set by locally aggregating neighbour solutions (e.g. average label count, cutoff calibration).
Evaluation compared multi-label classifiers (metric learners and transformers) combined with \rapt against global and label-wise thresholding baselines, and against few-shot LLMs. 
Across an industrial dataset and six public benchmarks, \rapt consistently outperformed global and label-wise static thresholding baselines. 
In the industrial setting, \rapt achieved its best predictive performance with metric learners, reaching 0.87 Macro-F1,
while fine-tuned transformer variants on average achieved 0.775 Macro-F1, 
outperforming few-shot LLM baselines ($K=5$) by $2\times$ and 
requiring at least $115\times$ less inference time and $13.5\times$ less GPU memory. 

\end{abstract}



\section{Introduction}

Unlike traditional single-label classification, multi-label classification requires predicting a set of labels for each instance. This capability has become central to many high-impact applications in text, vision, recommendation, and decision support~\cite{tsoumakas_multi-label_2007,zhang_review_2014}. 
The practical difficulty of multi-label classification is not only learning good label rankings, but deciding which labels to actually assign at test time when label cardinality varies across instances, and label co-occurrence patterns vary across the instance space. 
Typically, models output per label activations that imply a label ranking, but deployment requires a cutoff rule that maps activation scores to a variable size label set, often with unequal penalties for false positives (FPs, suprious labels) and false negatives (FNs, missed labels). 
Global decision rules, such as fixed score thresholds, are attractive for simplicity. Still, they can be hard to specify in skewed spaces where the semantics of a given score value can differ across regions of the instance space and across labels.

\begin{figure}[htb]
    \centering
    \includegraphics[width=1.0\textwidth]{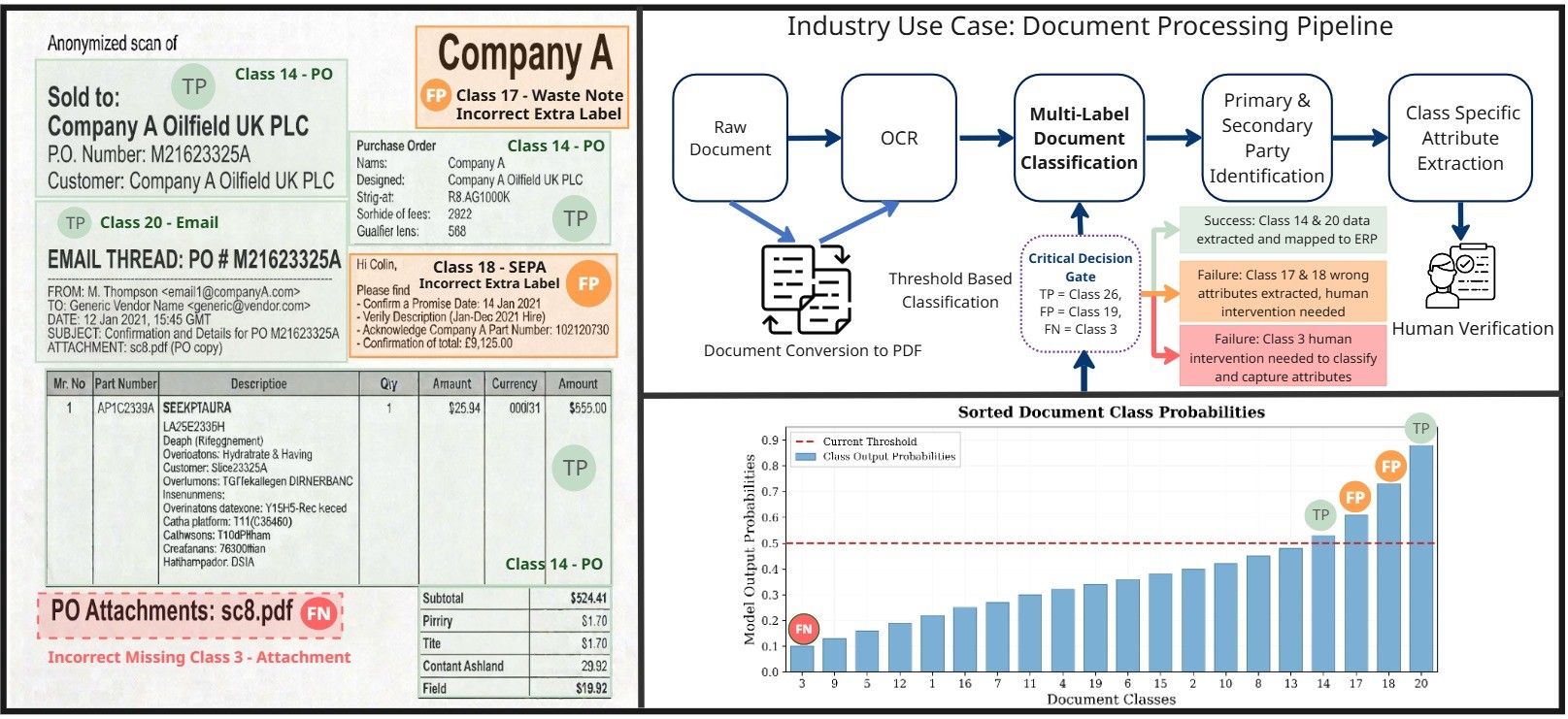}
    \caption{Industry use case pipeline for document processing and information extraction. Left: a typical document with areas indicated for class label assignment. Top: multi-label predictions act as routing decisions for downstream extraction tasks. Bottom: typical prediction scores from a backbone model on y-axis, shown relative to class specific optimal thresholds, with x-axis showing the label set. This illustrates how a global threshold (set at 0.5 for instance) can misclassify labels, omitting required classes (class label 3)  and including spurious ones (class label 17 and 18).}
    \label{fig:pipeline}
\end{figure}

In document understanding pipelines, multi-label predictions often determine pre-processing pathways.
A typical workflow, shown in Fig.~\ref{fig:pipeline}, converts raw documents to PDF, applies OCR, performs document categorisation, identifies areas of interest, and then runs document type and class specific attribute extraction. 
In Fig.~\ref{fig:pipeline}, the predicted label set triggers downstream tasks, so selection errors can propagate through subsequent stages before human verification.
Omitted labels (FNs) may prevent appropriate tasks from running, while spurious labels (FPs) can trigger inappropriate processes, yielding irrelevant or misleading fields and increasing verification effort and operational risk. 

This challenge is particularly evident in industrial settings, where organisations maintain large collections of legacy documents in heterogeneous formats, including scanned paper records, reports, and correspondence containing operational and customer information. 
In collaboration with \companyname, an industry partner operating in the transportation and waste management domain, we consider document processing pipelines that must handle diverse document types whose content may correspond to multiple operational categories simultaneously.
Accurate prediction of the label set is therefore critical to routing each document correctly to downstream extraction, review, and case handling workflows.

To address this multi-label classification challenge, we introduce \rapt, a lightweight retrieval-augmented wrapper employing case-based reasoning (CBR).
It can be applied post-hoc to any multi-label classifier producing embeddings and label-wise scores, enabling instance-specific thresholding and label set selection. In addition to benchmark data, the dataset used in this study is derived from \companyname's operational documents which provides a realistic setting where thresholding decisions directly influence downstream processing. The contributions of this work are outlined as follows:
\begin{itemize}
\item propose a framework for constructing a case base from the training set, in which each case consists of a multi-label classification instance and threshold values inferred from the ground truth labels associated with its score vector, eliminating the need for additional data or human annotation; 
\item introduce neighbourhood-driven dynamic thresholding strategies for variable cardinality set prediction, including AvgCount and Rank Calibration, and show how residual error information from retrieved cases can be used to adapt the multi-label score vector prior to thresholding; and
\item evaluate \rapt on an industrial document workflow dataset, and six public benchmarks showing consistent gains over global and label-wise thresholding baselines, with zero-shot and few-shot LLM comparators also reported. 
\end{itemize}

\section{Related Work}
\label{sec:related}

\textbf{Multi-label classification methods} are commonly divided into problem transformation and algorithm adaptation approaches~\cite{tsoumakas_multi-label_2007,zhang_review_2014}. 
Methods such as Binary Relevance, Classifier Chains~\cite{read_classifier_2011}, RAkEL~\cite{kok_random_2007}, and Calibrated Label Ranking~\cite{furnkranz_multilabel_2008} differ in how they model label dependence, but they still typically produce label-wise scores or rankings that must be converted into a final predicted label set. 
A central but often underemphasised part of this conversion is therefore thresholding: deciding where the relevance boundary should lie for a given instance. 
Prior work has studied \textbf{global and label-specific} thresholding strategies, including SCut~\cite{yang_study_2001}, MCut~\cite{hutchison_mcut_2012}, MetaLabeler~\cite{tang_large_2009}, MALTOR~\cite{jiang_relevance_2022}, and more recent neighbourhood-aware approaches~\cite{shamatrin_adaptive_2025}.
However, most methods either learn global thresholding rules or rely on additional predictive models, leaving open the question of how thresholding decisions might instead be derived directly from local evidence in similar labelled instances.


\textbf{Case-based reasoning} solves a new problem by \textbf{retrieving} similar prior cases and \textbf{adapting} their solutions to the current query~\cite{aamodt_case-based_1994}. 
This distinguishes CBR from neighbourhood-based methods that primarily aggregate local evidence, such as ML-kNN~\cite{zhang_ml-knn_2007} or Deep k-Nearest Neighbours~\cite{papernot_deep_2018}. 
In this sense, useful retrieval is not only about finding similar neighbours, but about whether their solutions can be revised for the new instance~\cite{smyth_adaptation-guided_1998,craw_learning_2006}. 
Our work adopts this adaptation perspective, but applies it to a different target: rather than using retrieved cases for direct label voting, we use them to adapt the score-to-set conversion step in multi-label prediction.

\textbf{Episodic metric meta-learning} provides a natural backbone for our framework because it jointly supports retrieval and prediction.
\textbf{Matching Networks} ~\cite{vinyals_matching_2017} condition predictions on a labelled support set, \textbf{Prototypical Networks}~\cite{snell_prototypical_2017} summarise classes by embedding prototypes, \textbf{Relation Networks}~\cite{sung_learning_2018} learn a comparison function, and \textbf{SNAIL}~\cite{mishra_simple_2018} combines temporal convolutions with attention for few-shot adaptation.
These models are especially attractive here because they produce embeddings suitable for similarity return (neighbour retrieval) while also yielding per-label scores that can be post-processed by our case-based adaptation rules.
Multi-label meta-learning remains comparatively underexplored.
Simon et~al.~\cite{simon_meta-learning_2021} introduced a few-shot multi-label meta-learning framework with an explicit label-counting component, which is directly relevant to dynamic thresholding.
Wu et~al.~\cite{wu_learning_2019} studied learning policies for multi-label heterogeneity.
Our framework is intentionally broader than any single meta-learning architecture: any backbone that provides an embedding space and per-label confidence scores can be paired with our retrieval-based adaptation layer.

\textbf{Fine-tuned Transformers} are a practically important family of backbones formed by transformer encoders fine-tuned for multi-label classification.
Representative models include \textbf{BERT-base} and \textbf{BERT-large}~\cite{devlin_bert_2019}, \textbf{DeBERTa-v3-base} and \textbf{DeBERTa-v3-large}~\cite{he_debertav3_2021}, as well as domain-specialised encoders such as LegalBERT~\cite{chalkidis_legal-bert_2020}, ClinicalBERT~\cite{alsentzer_publicly_2019}, and PubMedBERT~\cite{gu_domain-specific_2022}.
In these models, the \texttt{[CLS]} representation provides a compact retrieval embedding, while sigmoid-activated logits provide per-label scores for downstream thresholding and adaptation.
This combination makes \textbf{fine-tuned transformers} especially convenient for our framework: they are strong discriminative classifiers, but they also expose an embedding for case representation on which case retrieval can be built.

\textbf{Large language models} broaden the space of possible backbones and teachers.
Open-weight LLM families such as Llama~\cite{touvron_llama_2023}, OpenAI's gpt-oss models~\cite{openai_gpt-oss-120b_2025} and Qwen \cite{yang_qwen3_2025} make it increasingly feasible to use larger generative models for classification, retrieval, synthetic data generation, or weak supervision.
However, LLMs also introduce practical complications for our setting.
Compared with compact fine-tuned encoders, they typically incur higher inference latency, larger memory footprints, and greater serving cost; retrieval behaviour may also depend more strongly on prompt format, instruction tuning, and context budget ~\cite{zhou_survey_2024}.
For a framework that performs case retrieval at prediction time, these considerations are not incidental: they affect whether retrieval remains lightweight enough to be useful in practice.
For this reason, we view LLMs primarily as powerful comparative baselines, whereas compact fine-tuned encoders and meta-learners remain especially well matched to case-based test-time adaptation and in our work as the basis for the thresholding CBR wrapper.

\section{CBR Wrapper Methodology for Thresholding}
\label{sec:methodology}

We formulate multi-label thresholding as a CBR problem implemented as a wrapper
around an existing classifier. Rather than modifying the base model, the wrapper reuses the model's
score outputs and internal representations to construct a case base from labelled instances.
Given a query instance, the method retrieves similar prior cases (forming a neighbourhood) and adapts their thresholding
evidence to produce instance-specific label set predictions for the query. 
This forms a CBR cycle in which the
query defines the problem, retrieved labelled instances provide prior cases, with weighted aggregations forming case reuse and threshold
adaptation corresponds to the revise step.
If desired, newly verified instances may be retained as new cases, allowing the case base to expand over time, although this retain step is not the focus of the present paper.
\begin{figure}[htb]
\centering
\includegraphics[width=1.0\textwidth]{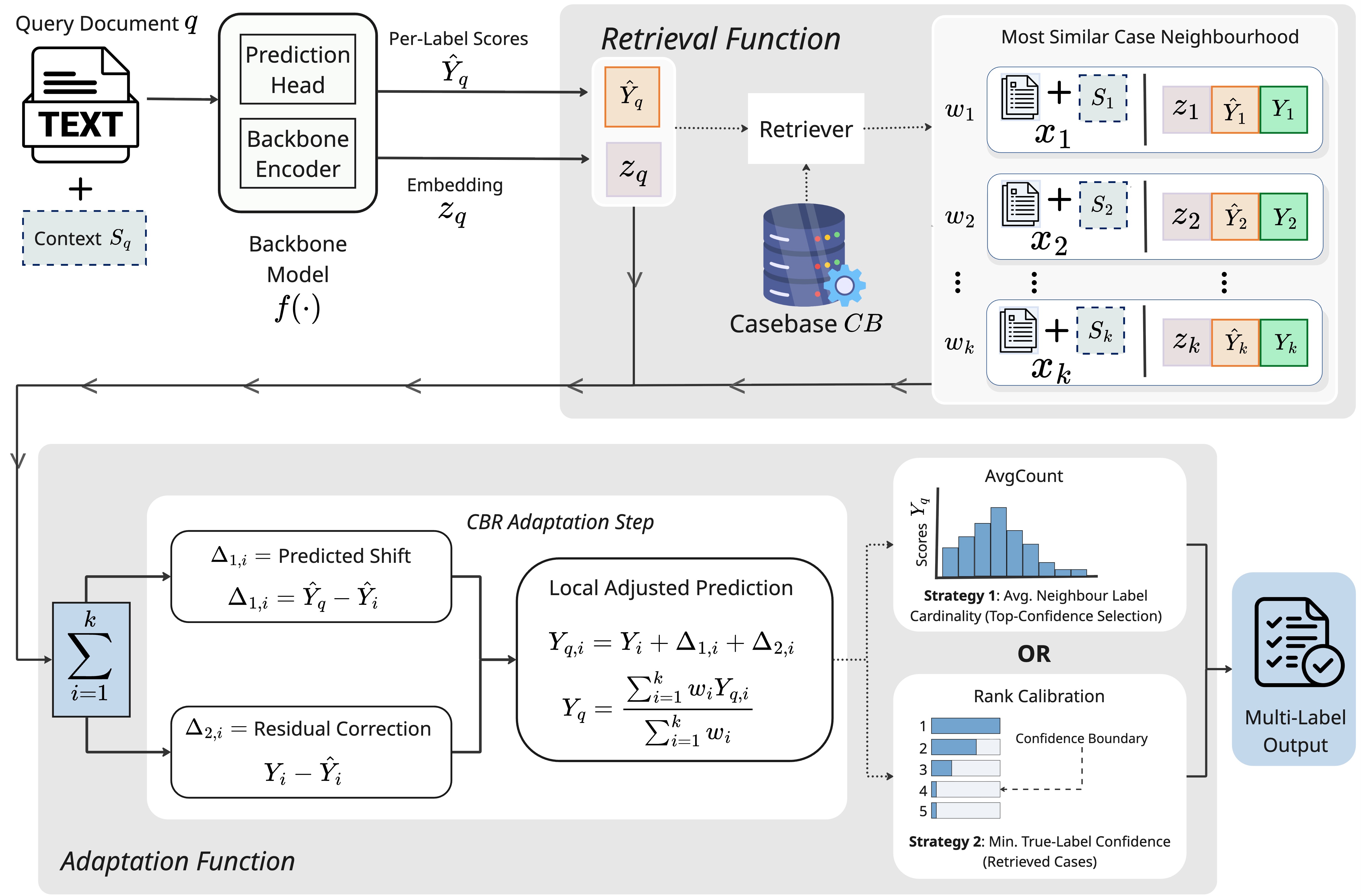}
\caption{CBR wrapper components, illustrated against a grey background. Given a query document, the backbone model, $f$, produces label scores and an embedding, which are used to retrieve similar cases from the casebase. Retrieved predictions and labels are then combined in an adaptation step to produce a locally adjusted prediction, followed by threshold calibration to obtain the final multi label output.
}
\label{fig:example}
\end{figure}




\subsection{Thresholding Case Base Construction}

The case base is constructed from existing training (and if available any validation data), requiring no additional data collection or annotation. 
It can be updated incrementally as newly verified documents become available, making the approach easy to integrate into existing document workflows without retraining.

Let $f$ denote a trained multi-label classification model over a label space of size $L$ that produces label-wise scores. For an input instance $x$, optionally accompanied by additional context $S$, the model produces a score vector $\hat{Y} \in \mathbb{R}^L$ and an associated representation $z \in \mathbb{R}^d$, written as $(z,\hat{Y}) = f(x,S)$. 
The vector $\hat{Y}$ contains the label-wise scores produced by the model's multi-label classification head, which may correspond to logits, activation values, or calibrated probabilities depending on the underlying architecture. Here $S$ captures model-specific context when available, such as support sets in metric-based learners, and is null otherwise.
We denote by $g(\cdot)$ the embedding function associated with $f$, such that $z = g(x,S)$. In practice, $g$ may be instantiated by an internal representation of the model, such as a final hidden layer embedding.
Using each labelled training instance $(x,Y)$, we construct a case
\begin{equation*}
\mathcal{C} = \langle x,\; S,\; z,\; \hat{Y},\; Y \rangle
\end{equation*}
where $Y \in \{0,1\}^L$ 
denotes the ground truth binary label vector.
Since the ground truth labels, $Y$, are known in the training data, and the model provides the $\hat{Y}$ prediction scores, thresholding evidence can be derived automatically without additional data or human annotation.
In particular, when $Y_{\ell}=0$, the case indicates that label 
$\ell$ is absent and therefore contributes no thresholding evidence for that label. 
The set of all such constructed cases forms the case base $\mathcal{CB}$.

\subsection{Case Retrieval and Similarity Weighting}

Given a query instance $q$ with optional context $S_q$, 
we retrieve the $k$ most similar cases from the case base $\mathcal{CB}$ as follows:

\begin{equation*}
\operatorname{Retrieve}(q,\mathcal{CB})
=
\operatorname*{TopK}_{\mathcal{C}_i \in \mathcal{CB}}
\left( w(z_q, z_i) \right)
=
\{\mathcal{C}_1,\dots,\mathcal{C}_k\}
\end{equation*}
where $w(\cdot,\cdot)$ denotes cosine similarity.
Here the query embedding is
$z_q = g(q, S_q)$ and $S_q$ is optional.
The similarity score of each retrieved neighbour is also used as its importance weight,
denoted $w_i$.

Given the retrieved cases, the goal is to reuse their label-wise thresholding evidence for the query instance. 
Since the query is rarely identical to any retrieved case, this reuse must be accompanied by an adaptation step that adjusts the retrieved thresholding evidence to the local characteristics of the query before final label set prediction.

\subsection{Local Neighbourhood Adaptation Step}

Given the retrieved neighbourhood $\{\mathcal{C}_1,\dots,\mathcal{C}_k\}$, adaptation is performed in score space by transferring case-specific evidence from the retrieved cases to the query.
The core idea is to begin from the query's own model output and then refine it using the observed prediction residuals of similar past cases. In this way, adaptation preserves the model's instance-specific prediction while incorporating corrective evidence from the local neighbourhood.
\begin{equation*}
Y_q = \operatorname{Adapt}(\hat{Y_q},\{\mathcal{C}_1,\dots,\mathcal{C}_k\}).
\end{equation*}

The locally adapted score vector of the query is  inferred from each neighbouring case, $\mathcal{C}_i$'s predicted score vector, $\hat{Y_i}$, as follows:
\begin{equation*}
Y_{q,i} = \hat{Y}_i + \Delta_{1,i} + \Delta_{2,i}.
\end{equation*}
Basically each retrieved case contributes a residual-based correction to the query's original score vector.
Here the two $\Delta$ terms are derived from each retrieved case $\mathcal{C}_i$ by using: 
1) a \textbf{Model Adaptation ($\Delta_{1,i}$)} term, which is the predicted score vector difference between the query and the retrieved case,
$\Delta_{1,i} = \hat{Y}_q - \hat{Y}_i$.
This term aligns the retrieved case with the query by shifting the neighbour's score profile toward the query's own model output; and 
2)
a \textbf{Residual Correction} ($\Delta_{2,i}$), which is a case-specific correction derived from the discrepancy between the retrieved case's ground truth labels and its score vector, $\Delta_{2,i} = Y_i - \hat{Y}_i$.
Here $\Delta_{2,i}$ captures how the model erred on the retrieved case, and transfers that error pattern as corrective evidence for the query.

The final adapted query score vector is obtained by similarity-weighted aggregation over the retrieved neighbourhood:
\begin{equation*}
Y_q = \frac{\sum_{i=1}^k w_i Y_{q,i}}{\sum_{i=1}^k w_i}.
\end{equation*}
In this manner, neighbours that are more similar to the query contribute more strongly to the final adapted score estimate.
The resulting vector $Y_q$ is a neighbourhood informed score estimate which is then converted into the final binary label set through dynamic thresholding.

\subsection{Inference and Dynamic Thresholding}
The final multi-label prediction is obtained from the adapted score vector $Y_q$ via dynamic thresholding, with the indicator function applied element-wise:
\begin{equation*}
Y_{\text{final}} = \mathbbm{1}(Y_q > \tau_q)
\end{equation*} 

The query specific threshold $\tau_q$ can then be determined via two alternative neighbourhood strategies.
Under \textbf{AvgCount}, we estimate the expected number of active labels for the query by computing a similarity-weighted average of the neighbours' label cardinalities:
\begin{equation}
k^* = \frac{\sum_{i=1}^k w_i \lVert Y_i \rVert_1}{\sum_{i=1}^k w_i}.
\label{eq:count}
\end{equation}
We then set $\tau_{\mathrm{Count}}$ so that thresholding $Y_q$ yields approximately $k^*$ positive labels (equivalently, we threshold between the $\lfloor k^* \rfloor$-th and $(\lfloor k^* \rfloor+1)$-th largest entries of $Y_q$).
The intuition is that the neighbourhood provides a local estimate of how many labels are typically active for instances in this region of the space.
Under \textbf{Rank Calibration}, the threshold is set from the weighted average of the minimum confidence scores assigned to positive labels in the retrieved neighbourhood:
\begin{equation}
\tau_{RCal} = \frac{\sum w_i \cdot \min \{ \hat{Y}_i^{(l)} \mid Y_i^{(l)} = 1 \}}{\sum w_i}
\label{eq:rank}
\end{equation}
this uses, for each neighbour, the lowest score that still corresponds to a relevant label as a proxy decision boundary. Averaging these boundaries produces a threshold that reflects how far scores typically need to rise before a label is considered positive in similar cases (i.e. locality thresholding knowledge).

\section{Evaluation}
\label{sec:evaluation}

The evaluation assesses whether the proposed \rapt thresholding wrapper improves label set selection in realistic document understanding settings. We compare five thresholding strategies:
\begin{itemize}
\item \textbf{\bhalf}: a fixed global threshold where $\tau = 0.5$ applied in
$Y_{\text{final}} = \mathbbm{1}(Y_q > \tau)$;
    
\item \textbf{\bopt}: a globally optimised threshold selected on the validation set using a greedy search procedure to maximise F1;

\item \textbf{\bloc}: a label-wise optimised threshold, $\tau_\ell$ optimised, on the validation set, where each label receives its own threshold through greedy tuning;
    
\item \textbf{\cav}: CBR wrapper thresholding methods using dynamic $\tau$ from AvgCount, (see Eq.~\ref{eq:count}); and

\item \textbf{\crk}: CBR wrapper using $\tau$ from rank calibration (RankCal), (see Eq.~\ref{eq:rank}).
\end{itemize}

The validation tuned baselines, \bopt and \bloc, provide stronger comparisons than the naive baseline \bhalf, as they optimise thresholds to each dataset. This allows us to test whether wrappers improve not only naive thresholding, but also competitive static thresholding procedures derived from held-out validation data.
For the proposed CBR thresholding wrapper, we set the retrieval neighbourhood size to $k=10$ by default. We also control the number of unique retrieved query situations retained during adaptation, denoted $n$, and set this to $3$ by default. Both parameters can be varied, and ablation studies examining sensitivity to these choices are reported in the supplementary material.

\subsection{Backbone models for multi-label classification}
We evaluate all thresholding methods with two main backbone families instantiating the scoring function 
$f(\cdot)$ in Fig.~\ref{fig:example}: 
episodic metric learners and transformer encoders. 
For metric learners, we consider Matching Networks, Prototypical Networks, Relation Networks, and SNAIL. 
For transformers, we evaluate the pretrained, head-only fine-tuned, and fully fine-tuned variants 
including domain-specialised encoders where appropriate.
We additionally report large language model (LLM) baselines using Qwen3-32B as a comparative reference, given their growing popularity as prompt-based alternatives for document understanding. 
While not part of the proposed deployment pipeline, they serve as important industrial comparators for assessing predictive effectiveness, inference efficiency, and serving cost relative to the proposed wrapper.
For episodic metric learners, evaluation is restricted to the labels present in the support set for a given episode, since these models only produce predictions over classes observed in support. All thresholding methods applied to metric learners are therefore compared under the same support-constrained setting. Transformer models, by contrast, output scores over the full retained label space and are evaluated across all labels. The same full-label evaluation setting is used for the LLM baselines; prompts are provided in the appendix.

\subsection{Datasets}
While our main focus is the \companyname industrial dataset, we also evaluate \rapt on six public multi-label benchmarks to assess its generalisation beyond this application setting: 
news (Reuters-21578 ~\cite{lewis_reuters-21578_1997}), 
legal (EUR-Lex ~\cite{chalkidis_large-scale_2019}), 
Event Classification ~\cite{vasylevskyi_events_classification_biotech_2025}, 
medical (MIMIC-III ~\cite{johnson_mimic-iii_2016}, OHSUMED ~\cite{croft_ohsumed_1994}), 
and e-commerce (AmazonCat-13K ~\cite{jain_extreme_2016}).
Together, these datasets support complementary evaluation goals: \companyname assesses practical utility in a document workflow setting, while the public benchmarks assess robustness and transferability across broader domains and datasets (Table~\ref{tab:datasets}).
To keep experiments computationally manageable, we restrict each dataset to its 100 most frequent labels (50 for MIMIC-III), providing a consistent setting across industrial and public data while preserving key multi-label challenges: imbalance, variable label cardinality, and threshold selection.

\begin{table}[htb]
\caption{Evaluation datasets overview. For datasets with more than 100 labels, we use the top 100 most frequent labels; MIMIC-III is restricted to the top 50 labels.}
\label{tab:datasets}
\centering
\small
\setlength{\tabcolsep}{6pt}
\renewcommand{\arraystretch}{1.2}
\begin{tabular}{l p{2.4cm} r r r r}
\toprule
Dataset & Domain & Train & Val & Test & \#Labels \\
\midrule
\companyname & Resource Mngt & 1,134 & 284 & 355 & 11 \\
Reuters-21578 & News & 6,182 & 1,546 & 3,005 & 100 \\
EUR-Lex & Legal & 47,996 & 12,000 & 5,000 & 100 \\
Event Classification & Events & 2,207 & 552 & 381 & 29 \\
MIMIC-III & Clinical & 33,560 & 8,390 & 7,404 & 50 \\
OHSUMED & Medical & 43,077 & 10,770 & 289,440 & 100 \\
AmazonCat-13K & E-commerce & 932,636 & 233,159 & 301,377 & 100 \\
\bottomrule
\end{tabular}
\end{table}
We report results using the original train/validation/test split for each dataset. 
Additional k-fold experiments examining robustness to partitioning effects are included in the supplementary material, so the main paper focuses on the standard split results, with particular attention to the industrial dataset.




\subsection{Evaluation Metrics and Implementation Details}
\label{sec:metrics_impl}

We compute standard multi-label metrics including Micro-F1, Macro-F1, Precision, Recall, and Hamming Loss.
For brevity, we report only Micro-F1 and Macro-F1, which together summarise overall performance and sensitivity to label imbalance; the remaining metrics are provided in the supplementary material.
We also assess deployment-oriented efficiency on the \companyname test set by measuring per-document inference latency and peak GPU memory usage, from which we derive hardware-normalised cost estimates (see Section~\ref{sec:inference_efficiency}).

All experiments were trained and evaluated on a single NVIDIA A100 80\,GB GPU.
This setup reflects the deployment-oriented framing of the work: the thresholding wrapper operates as a lightweight post-hoc layer on top of existing classifiers without requiring additional large-scale infrastructure.
All runs use a fixed random seed of 42 for reproducibility.

All 4 \textbf{metric learners} share the same episodic training protocol, sharing the same backbone encoder \texttt{sentence-transformers/all-mpnet-base-v2}.
Each episode samples $N\!=\!10$ classes, and $K\!=\!5$ support examples per class.
Strict document-level separation is enforced: the query document is never included in its own support set, preventing information leakage between query and support.
Training runs for up to 50 epochs with early stopping, using optimiser Adam (learning rate $10^{-4}$, weight decay $10^{-5}$) and batch size of 32.
All 4 models are trained per split under identical hyperparameters, ensuring a fair intra-family comparison. To ensure fair comparison across thresholding methods and with transformer baselines, the candidate label space at evaluation time is restricted to labels present in the support set of the current episode.


We evaluated 3 \textbf{Transformer} fine-tuning training regimes: (i)~\emph{pretrained}, where the model is used frozen without any task-specific training; (ii)~\emph{head-only}, where only a randomly initialised classification head is trained on top of the frozen encoder; and (iii)~\emph{fully fine-tuned}, where all parameters are updated end-to-end.
For regimes (ii) and (iii), we train for up to 25 epochs with early stopping (patience 5, improvement threshold $10^{-3}$).
The learning rate is $2 \times 10^{-5}$ for base-sized models and $1 \times 10^{-5}$ for large models, with linear warmup over the first 10\% of training steps and weight decay of $0.01$, with batch size 32.

\section{Results}


Our evaluation pursues three goals: assessing \rapt in its primary industrial application setting, examining whether its thresholding behaviour transfers to public multi-label classification dataset benchmarks, and comparing deployment efficiency against LLM baselines on the industrial task.

\subsection{\companyname Industry Domain Effectiveness Analysis}

Table~\ref{tab:wmnicol_main} reports Micro-F1 and Macro-F1 across the 4 metric learners and 4 transformer models on the baselines and \rapt
thresholding strategies.
Among the metric learners, \rapt performs best when combined with ProtoNet: \cav {} achieves a Macro-F1 of 0.8704, corresponding to relative gains of 10.8\% over \bhalf, 11.1\% over \bopt, and 2.8\% over \bloc. 
This is especially relevant in the presence of label imbalance, since Macro-F1 gives equal weight to each class and is therefore more sensitive to performance on rarer labels.

\begin{table}[htbp]
\centering
\caption{Micro \& Macro F1 across Model Configurations and Thresholding Methodologies. Method abbreviations: \bhalf~= Base ($\tau=0.5$), \bopt~= Base ($\tau=opt$), \bloc~= Base ($\tau=local$), \cav~= CBR + AvgCount, \crk~= CBR + RankCal.}
\label{tab:wmnicol_main}
\setlength{\tabcolsep}{3pt}
\scriptsize
\resizebox{\textwidth}{!}{%
\begin{tabular}{lcccccccccc}
\toprule
& \multicolumn{2}{c}{\bhalf} & \multicolumn{2}{c}{\bopt} & \multicolumn{2}{c}{\bloc} & \multicolumn{2}{c}{\cav} & \multicolumn{2}{c}{\crk} \\
\cmidrule(lr){2-3} \cmidrule(lr){4-5} \cmidrule(lr){6-7} \cmidrule(lr){8-9} \cmidrule(lr){10-11}
Models $f$ & Micro & Macro & Micro & Macro & Micro & Macro & Micro & Macro & Micro & Macro \\
\midrule
\multicolumn{11}{l}{\textbf{Metric Learners}} \\
MatchingNet & 0.9324 & 0.8441 & 0.9321 & 0.8111 & 0.9222 & 0.8354 & 0.9411 & \textbf{0.8607} & 0.9378 & 0.8502 \\
ProtoNet & 0.9283 & 0.7853 & 0.9301 & 0.7833 & 0.9378 & 0.8467 & 0.9467 & \textbf{0.8704} & 0.9426 & 0.8648 \\
RelationNet & 0.9332 & 0.8334 & 0.9147 & 0.7391 & 0.9357 & 0.8600 & 0.9424 & \textbf{0.8636} & 0.9416 & 0.8587 \\
SNAIL & 0.8853 & 0.4696 & 0.8853 & 0.4696 & 0.7977 & 0.4583 & 0.9470 & \textbf{0.8694} & 0.9047 & 0.8146 \\
\midrule
\multicolumn{11}{l}{\textbf{Pretrained Models}} \\
bert-base & 0.2166 & 0.0963 & 0.2419 & 0.1853 & 0.2226 & 0.1941 & 0.7868 & \textbf{0.6312} & 0.7894 & 0.6273 \\
bert-large & 0.2438 & 0.1510 & 0.2506 & 0.1293 & 0.2562 & 0.2132 & 0.7664 & \textbf{0.5999} & 0.7740 & 0.5625 \\
deberta-v3-base & 0.2414 & 0.1437 & 0.2414 & 0.1437 & 0.2274 & 0.1885 & 0.7714 & \textbf{0.6149} & 0.7692 & 0.5987 \\
deberta-v3-large & 0.2304 & 0.1013 & 0.2304 & 0.1013 & 0.2093 & 0.1902 & 0.7456 & \textbf{0.5578} & 0.7411 & 0.5413 \\
\midrule
\multicolumn{11}{l}{\textbf{Head-Only Fine-tuned Models}} \\
bert-base & 0.0000 & 0.0000 & 0.4000 & 0.0571 & 0.3575 & 0.1802 & 0.8018 & 0.6312 & 0.7533 & \textbf{0.6596} \\
bert-large & 0.0043 & 0.0011 & 0.3995 & 0.0570 & 0.3374 & 0.1592 & 0.7834 & \textbf{0.5900} & 0.7297 & 0.5889 \\
deberta-v3-base & 0.0000 & 0.0000 & 0.3995 & 0.0570 & 0.2091 & 0.1901 & 0.7827 & 0.6287 & 0.7842 & \textbf{0.6387} \\
deberta-v3-large & 0.0000 & 0.0000 & 0.3995 & 0.0570 & 0.2159 & 0.2071 & 0.7285 & 0.5497 & 0.7333 & \textbf{0.5795} \\
\midrule
\multicolumn{11}{l}{\textbf{Fully Fine-tuned Models}} \\
bert-base & 0.8166 & 0.5944 & 0.8382 & 0.6950 & 0.8412 & 0.7688 & 0.8410 & \textbf{0.7753} & 0.8346 & 0.7639 \\
bert-large & 0.8154 & 0.6531 & 0.8318 & 0.7440 & 0.8342 & 0.7508 & 0.8357 & \textbf{0.7655} & 0.8215 & 0.7192 \\
deberta-v3-base & 0.8301 & 0.6973 & 0.8379 & 0.7076 & 0.8364 & 0.7492 & 0.8163 & \textbf{0.7628} & 0.8032 & 0.7433 \\
deberta-v3-large & 0.8418 & 0.7514 & 0.8515 & 0.7694 & 0.8565 & 0.7886 & 0.8512 & \textbf{0.7957} & 0.8404 & 0.7409 \\

\bottomrule
\end{tabular}
}
\end{table}

Transformer backbones are overall less effective than metric learners on this task, as none reach the Macro-F1 levels above 0.85 achieved by the strongest metric models.
Within the transformer family, however, the effect of \rapt depends strongly on the training regime.
Fully fine-tuned transformers show only modest gains over the strongest baseline, consistent with stronger score calibration; for example, DeBERTa-v3-large improves from 0.7886 (\bloc) to 0.7957 (\cav) in Macro-F1, a relative gain of 0.9\%.
However, pretrained and head-only models benefit substantially: pretrained BERT-base increases from Macro-F1\,=\,0.0963 (\bhalf) to 0.6312 (\cav), while head-only DeBERTa-v3-base increases from 0 to 0.6287.
These results suggest that \rapt is useful with transformers when applied to backbones with poorly calibrated scores, where it can recover useful multi-label performance without further fine-tuning. On the \companyname data, \rapt configurations achieved approximately $2\times$ the Macro-F1 of the best few-shot LLM baseline (Qwen3-32B, $K\!=\!5$) as seen in Table \ref{tab:wmnicol_llm}.

A close examination of misclassified documents in the industrial dataset revealed two recurring failure modes. 
The first is co-occurrence driven false positives. In 19 cases, \rapt introduced a spurious label that frequently co-occurred with the true label in the local neighbourhood. The clearest example was \emph{Hazardous Waste Note} being added to \emph{SEPA Note} or \emph{Waste Transfer Note} documents, with these label pairs showing roughly 150--180 co-occurrences in training; in practice, such documents are often bundled together as part of a single case. As a result, retrieved CBR neighbours often contain both labels, inflating the spurious class score above the selection threshold. \cav was particularly susceptible in this setting, whereas \crk partially mitigated the issue, recovering 9 of the 19 cases by applying a sufficiently conservative per-instance threshold.

The second failure mode is format--content ambiguity. In a small number of cases, Purchase Orders written in an email-like style (e.g., \emph{``Hi Alex, please use order number DCL~417\ldots''}) retrieved neighbours that were closer in format than in semantic content. This pushed the \emph{Purchase Order} score downward and the \emph{Email Correspondence} score upward. More broadly, this highlights that retrieval quality depends critically on the structure of the backbone embedding space: when embeddings overemphasise superficial formatting cues, retrieval-augmented thresholding can reinforce the wrong decision.

\begin{table}[htb]
\centering
\caption{LLM (Qwen3-32B) on \companyname dataset. Relative scores (\%) computed against the best ML and Transformer baselines.
Relative improvements are computed against the \\
$^\dagger$best ML baseline 
(ProtoNet+\cav, Micro-F1=0.9467, Macro-F1= 0.8704); and \\
$^\ddagger$Best transformer baseline: DeBERTa-v3-large (fully fine-tuned) + \cav (Micro-F1 = 0.8512, Macro-F1 = 0.7957). \\
$^\ast$ Semantic labels were provided and represent an upper-bound reference.
}
\label{tab:wmnicol_llm}
\small
\setlength{\tabcolsep}{6pt}
\renewcommand{\arraystretch}{1.15}
\resizebox{\textwidth}{!}{%
\begin{tabular}{lcc|cc|cc}
\toprule
& \multicolumn{2}{c}{F1 Score} & \multicolumn{2}{c}{Rel.\ Best ML$^\dagger$} & \multicolumn{2}{c}{Rel.\ Best Trans.$^\ddagger$} \\
\cmidrule(lr){2-3}
\cmidrule(lr){4-5}
\cmidrule(lr){6-7}

Prompting Strategy & Micro & Macro & Micro & Macro & Micro & Macro \\

\midrule
Zero-shot (Semantic)$^\ast$ & 0.6134 & \textbf{0.5220} & -35.2\% & -40.0\% & -27.9\% & -34.4\% \\
Zero-shot (Anonymized) & 0.0958 & 0.0845 & -89.9\% & -90.3\% & -88.7\% & -89.4\% \\
Few-shot $K=1$ (Anon.) & 0.0996 & 0.1057 & -89.5\% & -87.9\% & -88.3\% & -86.7\% \\
Few-shot $K=3$ (Anon.) & 0.4232 & 0.3648 & -55.3\% & -58.1\% & -50.3\% & -54.2\% \\
Few-shot $K=5$ (Anon.) & 0.4408 & 0.3745 & -53.4\% & -57.0\% & -48.2\% & -52.9\% \\

\bottomrule
\end{tabular}
}

\vspace{3pt}
\footnotesize
\end{table}

\subsection{\companyname Industry Domain Inference Efficiency Analysis}
\label{sec:inference_efficiency}

We profiled inference efficiency on the \companyname test set to assess deployment characteristics across all model families (Table~\ref{tab:efficiency}).
For each model, we measured per-document inference latency (mean and P95 via \texttt{time.perf\_counter()}), throughput (documents per second), and peak GPU memory (\texttt{torch.cuda}).
We additionally report GPU-hours per document as a hardware-agnostic cost proxy that practitioners can multiply by their accelerator's hourly rate.

From a deployment perspective, as expected metric-learners and transformer models combined with \rapt offer substantial advantages over LLM-based alternatives. Although GPU-time determines the deployment cost in our setup, prompt length strongly influences LLM inference latency.
Average token usage increased substantially with few-shot prompting, rising from approximately 1k tokens for zero-shot prompts to over 7.7k tokens for $K\!=\!5$ demonstrations.
This increase in prompt size explains the higher latency observed for larger few-shot versions and highlights the sensitivity of LLM inference efficiency to prompt construction.
\rapt configurations are at least $115\times$ faster (38--236\,ms vs.\ 27--48\,s per document) and required at least $13.5\times$ less GPU memory ($\leq$2.23\,GB vs.\ $>$30\,GB).
These characteristics make \rapt particularly suited to higher throughput (millions of documents) industrial document workflows.

\begin{table}[htbp]
\centering
\caption{Throughput is computed as $1000 / \text{mean latency (ms)}$. GPU-hrs/doc converts mean latency to GPU-hrs for cost estimation.}
\label{tab:efficiency}
\setlength{\tabcolsep}{3pt}
\scriptsize
\begin{tabular}{lccccc}
\toprule
Model & GPU (GB) & Mean (ms) & P95 (ms) & Tput (docs/s) & GPU-hrs/doc \\
\midrule
\multicolumn{6}{l}{\textbf{Metric Learners}} \\
MatchingNet & 0.47 & 105.31 & 120.25 & 9.50 & $2.93 \times 10^{-5}$ \\
ProtoNet    & 0.47 & 108.69 & 121.55 & 9.20 & $3.02 \times 10^{-5}$ \\
RelationNet & 0.47 & 110.47 & 125.05 & 9.05 & $3.07 \times 10^{-5}$ \\
SNAIL       & 0.47 & 235.78 & 297.80 & 4.24 & $6.55 \times 10^{-5}$ \\
\midrule
\multicolumn{6}{l}{\textbf{Transformers}} \\
bert-base        & 0.89 & 40.20 & 44.47 & 24.88 & $1.12 \times 10^{-5}$ \\
bert-large       & 1.78 & 93.17 & 99.72 & 10.73 & $2.59 \times 10^{-5}$ \\
deberta-v3-base  & 1.26 & 72.67 & 99.67 & 13.76 & $2.02 \times 10^{-5}$ \\
deberta-v3-large & 2.23 & 151.99 & 177.52 & 6.58 & $4.22 \times 10^{-5}$ \\
legal-bert-base  & 0.92 & 39.86 & 44.89 & 25.09 & $1.11 \times 10^{-5}$ \\
clinical-bert    & 0.89 & 37.56 & 42.83 & 26.62 & $1.04 \times 10^{-5}$ \\
pubmed-bert      & 0.89 & 39.91 & 45.72 & 25.06 & $1.11 \times 10^{-5}$ \\
\midrule
\multicolumn{6}{l}{\textbf{Large Language Models (Qwen3-32B)}} \\
Zero-shot (Semantic)$^\ast$ & 30.10 & 47661.59 & 74174.17 & 0.02 & $1.32 \times 10^{-2}$ \\
Zero-shot (Anon.)           & 30.10 & 28236.22 & 43396.21 & 0.04 & $7.84 \times 10^{-3}$ \\
Few-shot $K\!=\!1$          & 30.25 & 28344.47 & 50431.93 & 0.04 & $7.87 \times 10^{-3}$ \\
Few-shot $K\!=\!3$          & 30.55 & 27124.87 & 52785.33 & 0.04 & $7.53 \times 10^{-3}$ \\
Few-shot $K\!=\!5$          & 30.85 & 30397.81 & 53484.44 & 0.03 & $8.44 \times 10^{-3}$ \\
\bottomrule
\end{tabular}
\end{table}

\subsection{Public Benchmark Datasets Effectiveness Summary Analysis}
\label{sec:results_summary}
\begin{figure}[htb]
    \centering
    \includegraphics[width=\linewidth]{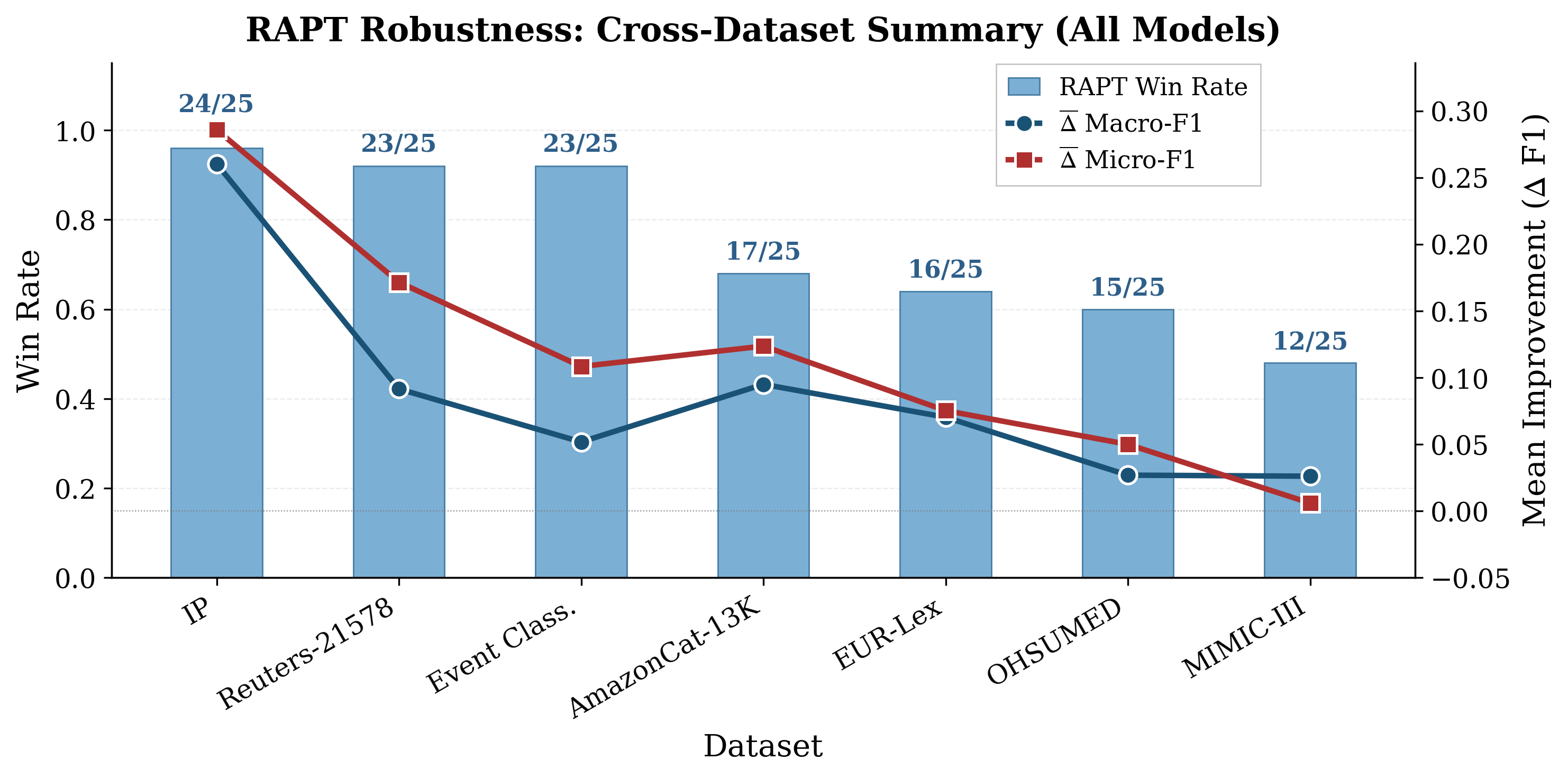}
    \caption{Cross-dataset accuracy summary across all models. Bars show the \rapt win rate per dataset, while lines show mean improvement in Macro-F1 and Micro-F1 relative to the best static baseline. Labels above bars indicate the number of wins out of 25 model configurations (4 metric learners + 7 Transformers x 3 modes).}
    \label{fig:cross_dataset_summary}
\end{figure}

For brevity, we summarise \rapt's effect across all seven datasets in Fig.~\ref{fig:cross_dataset_summary}, with full per-model results deferred to the Appendix.
Rather than comparing only the single best configuration per family, we report how often \rapt outperforms the best static baseline for \emph{each individual model}, along with the average F1 gain across all models.
Specifically, for each model we compare its highest Macro-F1 under any static threshold (\bhalf, \bopt, or \bloc) against its highest Macro-F1 under either \rapt variant (\cav or \crk).
Accordingly we have 175 model--dataset combinations, and \rapt achieves the higher Macro-F1 in 130 cases (74.3\%), with an average improvement of +0.093 Macro-F1 and +0.066 Micro-F1.
Among metric learners, win rates reach 75--100\% on every dataset, with average Macro-F1 gains of +0.06 to +0.14.

\section{Conclusion}
\label{sec:conclusion}

This paper addressed a practical problem in industrial multi-label document classification: how to convert model outputs (scores) into reliable document labels under noisy, imbalanced, heterogeneous document formats, and downstream workflow constraints.
Here, the labelling decision is a critical post-processing step, as it determines downstream extraction, routing, and review actions, directly affecting operational accuracy and verification effort.
 
To address this challenge, we introduced \rapt, a retrieval-augmented post-hoc thresholding wrapper that adapts label set selection at inference time using case-based reasoning (CBR).
Because \rapt requires only embeddings and label-wise scores, it can be paired with different backbone classifier model families, including metric learners and fine-tuned transformers, without retraining or architectural changes.
This makes it attractive for applied settings where organisations already use classifiers, but need a lightweight way to improve robustness as document collections evolve. Across 175 model--dataset combinations spanning an industrial document classification pipeline and six public benchmarks from legal, biomedical, clinical, news, event, and e-commerce domains, \rapt outperformed the best static threshold approach in 130 cases (74.3\%).
 
Looking ahead, we see two promising extensions.
First, the CBR \emph{retain} step could enable online case-base growth from newly verified documents, allowing the thresholding wrapper to adapt to distributional drift without retraining, which is valuable in settings where document templates and label distributions evolve over time.
Second, evaluating \rapt on extreme multi-label tasks with the full label space (e.g.\ AmazonCat-13K with 13{,}330 labels) would test whether neighbourhood-driven thresholding remains effective when label sparsity is severe.%
%
%
\bibliographystyle{splncs04}
\bibliography{references_shorten}

\end{document}